\documentclass[sigconf]{acmart}
\usepackage[inline]{enumitem}
\usepackage[version=4]{mhchem}

\usepackage{ccicons} 

\DeclareMathOperator{\logit}{logit}

\setcopyright{none}
\copyrightyear{}
\acmYear{}
\acmDOI{}

\acmConference[Preprint]{Preprint}{Decembre, 16, 2025}{}
\acmISBN{}

\usepackage{xspace}

\newcommand{\yl}[1]{\textcolor{black}{#1}}

\newcommand{\designers}{{\textit{\textbf{Designers}}}\xspace}

\newcommand{\reviewers}{{\textit{\textbf{Reviewers}}}\xspace}
\newcommand{\improvers}{{\textit{\textbf{Improvers}}}\xspace}
\newcommand{\IPCC}{{\textit{\textbf{IPCC Managers}}}\xspace}
\newcommand{\public}{{\textit{\textbf{Target Audience}}}\xspace}

\begin{document}

\title{A Constructive Scientific Methodology to Improve Climate Figures from IPCC} 

\author{Lu Ying}
\email{yiyinyingl@outlook.com}
\orcid{0000-0002-4206-231X}
\affiliation{%
  \institution{Zhejiang University \& Inria}
  \country{China, France}
}

\author{Junxiu Tang}
\email{junxiu.tang@kellogg.northwestern.edu}
\orcid{0000-0003-3594-926X}
\affiliation{%
  \institution{Northwestern University}
  \country{USA}
}

\author{Tingying He}
\email{hetingying.hty@gmail.com}
\orcid{0000-0002-9670-5587}
\affiliation{%
  \institution{The University of Utah}
  \country{USA}
}

\author{Jean-Daniel Fekete}
\email{jean-daniel.fekete@inria.fr}
\orcid{0000-0003-3770-8726}
\affiliation{%
  \institution{Inria \& Université Paris-Saclay}
  \country{France}}

\renewcommand{\shortauthors}{Ying et al.}

\begin{abstract}
  We propose a methodology to improve figures from the Intergovernmental Panel on Climate Change (IPCC), ensuring that all modifications remain scientifically rigorous.
IPCC figures are notoriously difficult to understand, and although designers have proposed alternatives, these lack formal IPCC validation and can be dismissed by skeptics.
To address this gap, our approach starts from official IPCC figures. 
We gather their associated \emph{learning objectives} and devise \emph{tests} to score a pool of figure readers to assess how well they learn the objectives.
We define improvement as higher scores obtained by a comparable reader pool after viewing a revised figure, where all modifications undergo review to ensure scientific validity.
This assessment gives freedom to designers, who can deviate from the original design while making sure the objectives are still met and improved.
We demonstrate the methodology through a case study and describe unexpected challenges encountered during the process.
\end{abstract}


\begin{teaserfigure}
  \centering 
  \includegraphics[width=\linewidth]{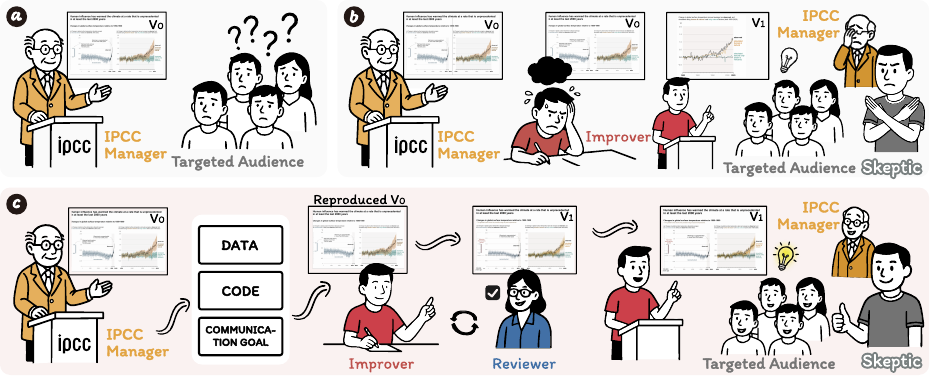}
  \caption{We compare the original IPCC figure creation process (a and b) with our proposed methodology for improvements (c).
(a) Without additional support, many audience often finds the figure difficult to understand.
(b) External improvements may raise skepticism.
(c) Our methodology establishes a chain of trust across distinct stakeholders (\IPCC, \reviewers, \improvers, and \public), ensuring the audience gains a clearer understanding of the figure and thereby achieves the intended communication goal. Black arrows illustrate a ``chain of trust'' connecting all stakeholders.}
  \label{fig:teaser}
\end{teaserfigure}


\maketitle

\section{Introduction}
The Intergovernmental Panel on Climate Change (IPCC) is the United Nations body responsible for assessing the science related to climate change~\cite{IPCC}. 
IPCC produces a range of reports, including comprehensive Assessment Reports, Special Reports on specific topics agreed to by its member governments, and Methodology Reports.
These reports have different versions catering to different audiences, mostly professional. 
This focus often means that the reports, though rich in valuable insights about the current state and future of the Earth's climate, are challenging to fully grasp for the general public. The specialized language and technical depth can make the content inaccessible to non-experts, when their understanding of climate change issues becomes more pressing every day.

A contributing factor to this challenge is the design of visualizations in these reports. While figures are generally considered the accessible and engaging components of a report~\cite{rodriguez2015improving}, the figures in IPCC reports are often created to serve the needs of scientists and policymakers, focusing on conveying complex data and insights with many details, using a language of specialists.
For example, the visualization figures in the \href{https://www.ipcc.ch/report/ar6/wg1/figures/summary-for-policymakers/}{Summary for Policymakers} in the \href{https://www.ipcc.ch/report/ar6/wg1/figures/}{AR6 WG1} often fail to effectively communicate to the public~\cite{schuster2024being}. 

As the official scientific source, IPCC figures carry strong potential for wider communication. Indeed, some external contributors, such as external media outlets, have begun to adapt them~\cite{Guardian2021}.
However, their approach also leads to increased skepticism. If changes to the presentation of data are made without stringent oversight, the validation of the information could be questioned. 
For example, if a newspaper publishes a modified version of a climate change figure, readers may question the reliability of this new version, whether it accurately represents the realities of climate change.
Such skepticism is beneficial in academic settings, as it fosters critical thinking and a deeper quest for factual knowledge. 
In less positive scenarios, critics could challenge the accuracy and legitimacy of revised figures, using these concerns to argue against actions aimed at mitigating global warming.

These challenges point to a broader gap: What would it take for external contributors to improve IPCC figures while preserving scientific fidelity, transparency, and trust?
This question guides our work. 
To answer it, we introduce a scientific improvement methodology that embeds a chain of trust to structure, document, and validate each refinement.
This approach is similar to established standards, such as the Graphic Reproducibility Stamp\footnote{\href{www.replicabilitystamp.org}{replicabilitystamp.org}}, which ensures that visual representations adhere to strict criteria for reproducibility and transparency.
We realize this chain of trust by structuring the improvement process around four key stakeholders:
\improvers to improve figures, 
\reviewers who should make sure the improved figures are faithful to the original design, 
\IPCC, which should provide communication goals and information associated with the figures,
and \public who receive figures that are more understandable and visually appealing.

We demonstrate our methodology through a case study with one figure from the IPCC Working Group 1. 
With our methodology, externals can begin by specifying the target audience for the figure improvement.
Improvers are responsible for making improvements in a transparent fashion.
Reviewers can validate the designers' improvements, and
IPCC climate scientists can facilitate the process by providing the specific information that needs to be conveyed to the audience.
Ultimately, the audience can receive figures that are more tailored to their needs.

The benefits of our methodology can extend beyond climate change and environmental organizations. For example, the Intergovernmental Science-Policy Platform on Biodiversity and Ecosystem Services (IPBES) met similar issues now~\cite{ipbes2019intergovernmental}. 
By following our methodology, more organizations can improve the transparency and impact of their data visualizations, ensuring complex information is communicated clearly and effectively to diverse audiences.

In summary, this work contributes to existing research by providing:
\begin{itemize}
    \item a constructive scientific methodology for improving IPCC figures, introducing a chain of trust among stakeholders,
    \item a case study demonstrating the application of the methodology to real IPCC figures,
    \item a description of the challenges encountered during the process and a discussion on the methodology.
\end{itemize}

\section{Related Work}

We report on work related to peer review, improving visualizations, and assessing their effectiveness. We also address transparency and reproducibility in visualization, which is essential for conveying information to a critical public.

\subsection{Peer Review}
Peer review is an essential mechanism for maintaining quality and integrity in scientific fields. 
In the context of scientific journal publications, peer review often involves a thorough evaluation by multiple experts who assess professional
performance, novelty, and quality of submitted work~\cite{lee2013bias}. 
Mayden highlighted the importance of peer review as ``golden standard''~\cite{mayden2012peer}. 
Tennant explored the current landscape of peer review in academia and scholarly communication, highlighting its roles in quality control, screening research, legitimizing scientific work, and enabling the self-regulation of scientific communities~\cite{tennant2018state}.
The rigorous nature of journal peer review ensures that only high-quality research is disseminated to the academic community.

In addition to traditional publication reviews, other forms of peer review have gained prominence, particularly in response to the growing emphasis on reproducibility and transparency in research. 
These forms are more focused and often involve a simpler, more objective evaluation process.
For instance, Peng discusses the peer review process for data and code, where reviewers evaluate whether the provided resources are sufficient and correctly implemented to reproduce the reported findings~\cite{peng2011reproducible}. 
The Open Materials Badge\footnote{\href{https://www.cos.io/initiatives/badges}{www.cos.io/initiatives/badges}}, introduced by the Open Science Collaboration, adds another layer to peer review by recognizing researchers who make their materials openly available, enabling reviewers to assess the completeness and usability of these materials for replication~\cite{kidwell2016badges}. 
In the computer graphics community, the Graphic Reproducibility Stamp Initiative (GRSI)\footnote{\href{www.replicabilitystamp.org}{replicabilitystamp.org}} involves a peer review process specifically designed to verify that graphical representations in research can be accurately reproduced by others, ensuring the reliability of mostly visual data.
The ACM acknowledges the reproducibility efforts made by its journals and conferences by providing three levels of reproducibility stamps associated with peer-reviewed publications. The actual process to obtain the stamps depends on the domain, journal, or conference, but it is becoming almost standard for articles with ``artifacts'' in conferences such as SIGMOD since 2022 (see the \href{https://reproducibility.sigmod.org/}{SIGMOD Reproducibility Availability \& Reproducibility Initiative}).

In our methodology, we also incorporate reviews into the IPCC figure improvement process to ensure the validation of improvements. 
Similar to badges and stamps, our approach is more objective and focused, aiming to validate various operations rather than conducting the comprehensive evaluations typical of traditional journal reviews. 
The key difference lies in the standards and participants involved. Our peer review process requires scientists to evaluate the work of \designers, focusing on scientific accuracy and effective communication, rather than the broader academic criteria used in traditional peer review. 
This ensures that the improved IPCC figures not only meet high scientific standards but also effectively convey complex information to a wide audience.

\subsection{Visualization Linting and Improvement}

To enhance the quality of visualizations, it is important to include both visualization linting and visualization improvement. 
Visualization linting is a concept first introduced by Meeks~\cite{meeks2017linting} and later systematically discussed by McNutt and Kindlmann~\cite{Mcnutt2018LintingFV}, which refers to automated evaluators that check visualizations against a set of predefined rules. Building on this idea, researchers developed many linters.
For example, McNutt et al. \cite{McNutt2020Surfacing} adapted the method of metamorphic testing to identify failures in charts.
Hopkins et al.~\cite{Hopkins2020visualLint} introduced VisuaLint, a linting technique that alerts users to possible errors in a visualization.
Chen et al.~\cite{VizLinter2022Chen} proposed a framework that includes a visualization linter and a visualization fixer to help users find and correct flaws.
In addition to general visualization linters, linting techniques for specific scenarios have also been proposed.
For example, Lei et al.~\cite{Lei2024GeoLinter} developed GeoLinter, a tool for linting choropleth maps, Hull et al. developed VisGrader~\cite{Hull2024VISGRADER} for grading D3 visualizations, and McNutt et al. developed \textsc{PaletteLint}~\cite{McNutt2025Mixing} for linting visualization color palette.

On the other hand, approaches aimed at improving visualization focus on making visualizations more usable in various conditions, such as scenarios involving large datasets or specific visual analytics requirements from different domains.
These improvements can target visual encoding or the layout of the visualization.
To achieve specific design goals, such as revealing data patterns~\cite{Palomo2016visually}, improving rendering efficiency~\cite{Zhu2021drgraph}, or enhancing perception~\cite{Micallef2017scatterplots}, novel algorithms are often proposed.
To better present a visualization, several studies proposed guidelines or methods to optimize the visualization content in dynamic contexts, such as data videos~\cite{Tang2020Guideline, Yao2024Swimflow}.

Our work is based on both visualization linting and improvements, as we need to detect chart flaws and improve charts during the iterative revision process of our methodology. However, our work further emphasizes the criteria of reproducibility and transparency for both linting and improvement.

\subsection{Visualization Assessment}
Evaluating whether a visualization communicates its intended message remains a core challenge in visualization research~\cite{Adar2021Communicative}. In scientific communication, where audiences often have limited prior knowledge, this challenge is amplified, making it crucial to use evaluation frameworks that directly link design choices to user understanding.

The information communication of visualizations depends on both the visualization itself and the readers. 
Prior work investigated assessment of various aspects of visualization, including perceived readability \cite{cabouat2025previs}, trust \cite{Pandey2023Trust,elhamdadi2024Vistrust}, aesthetics \cite{He2022Beauvis}, engagement \cite{mahyar2015towards, Hung2017Assessing}, etc. 
Researchers also developed several tools for evaluating the visual literacy of readers \cite{lee2017vlat,pandey2023minivlat, Ge2023CALVI, Cui2024Adaptive}.

Task-driven evaluation is another common approach, which measures performance on predefined tasks. These tasks range from low-level, such as perceiving values, identifying extremes, or finding clusters~\cite{cleveland1984graphical,amar2005lowlevel,quadri2021survey}, to high-level objectives, such as generating insights or new knowledge~\cite{quadri2024you,yalcin2016cognitive, Saraiya2005insight}. 

However, task-driven evaluations have limitations in aligning design intent with user outcomes. 
Because tasks are predefined by evaluators, they do not reflect how people encounter visualizations in the wild, where no explicit task is given. 
Prior work shows that, when viewing real-world charts, only 41\% of participant descriptions fully matched the designer’s intended message~\cite{quadri2024you}, revealing a substantial design–outcome gap.
To address this gap, Robbins et al.~\cite{Robbins2022Learning,lee-robbins2023affective, Adar2021Communicative} proposed using explicit learning objectives—statements of what a viewer should understand or be able to do after viewing—to align design goals with evaluation. Such objectives, grounded in frameworks like Bloom’s Taxonomy~\cite{krathwohl2002revision}, extend evaluation beyond perception to higher-level comprehension and can be directly translated into measurable tasks for communicative visualizations.
In this work, we adapt the learning-objective framework to guide the improvement of IPCC figures and evaluate how successfully we achieve the communication goals of the figures.

\subsection{Replicability and Transparency in Visualization}

Reproducibility and replicability are critical concerns in research, particularly when data is involved. 
Freire et al.~\cite{freire2016reproducibility} emphasize its importance within data-driven experiments across diverse scientific domains, and Koukouraki and Kray~\cite{koukouraki2023map} highlight that clear documentation and standardized practices are essential for ensuring reproducibility in geoscientific data preparation and presentation. 
In the domain of data visualization, Fekete and Freire~\cite{fekete2020Reproductibility} discuss the methodological challenges of achieving reproducibility, showing that even seemingly minor choices in encoding or rendering can alter outcomes.

Beyond reproducibility, transparency plays a complementary role by enabling verification and fostering trust in research outputs.
These values resonate with the FAIR Guiding Principles for Scientific Data Management and Stewardship, which emphasize making data Findable, Accessible, Interoperable, and Reusable to support openness and accountability in science~\cite{wilkinson2016fair}.
Prior work in visualization has proposed various strategies to improve transparency, often by making additional information available to end-users. 
For example, providing metadata such as data provenance can help signal credibility and build trust~\cite{burns2024invisible, dork2013critical}. 
More recently, researchers have explored the use of badges displayed alongside visualizations to communicate data sources and design considerations, thereby enhancing transparency~\cite{edelsbrunner2025visualization}.
Together, reproducibility and transparency form the foundation of credible and trustworthy scientific communication.

Since its Sixth Assessment Report (AR6), IPCC has published its data and most of its figures in open source format for reproducibility~\cite{Bush2020Perspectives}, in addition to publishing them in open access.
The assumption is that, if someone wants to make sense of the rationale behind IPCC reasoning, they can track the science through the publication, get the climate data collected, recompute the models to check the conclusions, and reuse the visualizations to better understand the data. If information is not accessible, the process is not replicable~\cite{Bush2020Perspectives}, and a critical scientist can, in good faith, question the report's conclusions.

Within the IPCC analysis and publication chain, the visualization part plays a particular role since it can be tailored to answer questions for specialists and the general public, starting from the same data.
However, currently, IPCC only produces visualizations tailored to specialists, hardly reproducible~\cite{ying:hal-04744236}. 
Someone else can take the responsibility to produce visualizations for non-specialists, but they need to start from actual data to convey the right information and the information correctly.
This gap highlights the need for systematic and transparent methodologies to guide figure improvements, ensuring both scientific rigor and accessibility for broader audiences.
\section{IPCC Process for Figures}
\label{sec:background}

This section introduces how the process of IPCC figure revision currently works and how it inspires our framework design.
We discussed with a data scientist from the IPCC technical support unit, who oversees the data management for \href{https://www.ipcc.ch/working-group/wg1/}{Working Group I}. They provided insight into the current figure improvement process for IPCC reports. 

In the full report, each chapter corresponds to a specific scientific field. For example, \href{https://www.ipcc.ch/report/ar6/wg1/chapter/chapter-8/}{Chapter 8} covers meteorology, hydrology, and atmospheric sciences, and \href{https://www.ipcc.ch/report/ar6/wg1/chapter/chapter-9/}{Chapter 9} focuses on oceanography and cryospheric science. 
The figures in these chapters in the full report are typically based on already published research, so they require minimal revision. 
However, the situation differs for the summary reports for policymakers and technical reports. 
For the summary reports, the figures are determined through a collaborative process where climate scientists gather to negotiate revisions. Because of the need for consensus across diverse expertise and countries, this process is usually lengthy and challenging. This difficulty is amplified when the figures are infographics. These infographics are central for communicating complex systems, but are not directly backed by datasets.
Such visuals often require particularly careful simplification and redesign.

Our proposed framework aims to simplify this revision process by organizing the collaboration involved in figure improvement.
To do so, we carefully examined how IPCC figures move through their existing production and communication workflow.
In practice, a figure typically passes through several groups of people: 
climate scientists who bring their domain expertise to define the scientific meaning of a figure; groups of scientists and government representatives who discuss and negotiate revisions to ensure that the scientific messages remain intact; contributors who propose visual or communicative improvements; and, ultimately, the audiences for whom these improved figures are intended.
Each of these groups plays a distinct role in shaping how a figure is created, validated, and understood.
Recognizing these naturally occurring divisions of responsibility provides the foundation for our framework, which formalizes these roles and clarifies how they interact during the improvement process.
Building on these observations, we organize these responsibilities into four stakeholder roles, which were briefly introduced earlier and are detailed below.

\begin{itemize}
    \item \improvers: this group comprises individuals motivated to enhance climate change figures and equipped with the skills to do so. Improvers possess the ability to both conceptualize and execute design changes. They may include designers, and other citizens aiming to communicate climate change concepts effectively to non-specialists, such as university students. 
    \item \reviewers: this group consists of two types of experts: climate scientists and visualization experts. Climate scientists are the primary reviewers, but visualization experts are consulted when climate scientists encounter difficulties understanding the process or encodings for generating the figure.
    \item \IPCC: these individuals are part of the IPCC and are responsible for defining and providing the communication goals of the IPCC figures, ideally in the form of learning objectives and corresponding assessments. They should also provide the source code and data needed to generate accurate and reliable figures. They play a crucial role in guiding the overall process and ensuring that the figures align with IPCC standards and objectives.
    \item \public: this group consists of the individuals who will ultimately engage with the improved figures. The target audience can vary depending on the context. For example, university students may improve figures for their peers, and company employees may enhance figures for their colleagues.
\end{itemize}

\section{Methodology for Improving IPCC Figures}
This section outlines our approach to improving visualization figures in IPCC reports, ensuring high-quality, accurate, and transparent improvements while maintaining fidelity to the original data.
We begin with an overview of the entire methodology, followed by detailed explanations.

\subsection{Overview}
Our methodology follows an iterative design process.
The idea consists of (1) constraining \improvers to produce an improved visualization by using an existing IPCC report figure as the starting point and (2) iteratively refining it to adhere to a final design that performs better on the comprehension test, yet stays faithful to the original data and intentions.
The process begins from an initial state (\autoref{sec:initial_state}) and proceeds through several iterations. Details of each iteration are described in \autoref{sec:iteration}.

The iterations continue until the final design is successful.
A test designed from \emph{learning objectives}, such as a questionnaire with a specific \emph{scoring rule}, is used to evaluate the final design. 
Multiple conditions can determine the ending point of the iterations. 
A classic condition is reaching a predefined threshold based on the scoring rule. 
Frequently, in industry as well as research, resource constraints, such as time and budget limits, can define the endpoint. 
We consider that the new design is an improvement compared to the baseline figure only if the score is improved by the scoring rule at the endpoint, at least for a specified population segment. 

\begin{figure*}
    \centering
    \includegraphics[width=\linewidth]{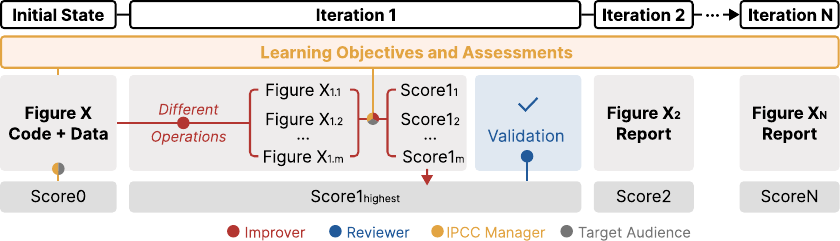}
    \caption{The methodology for improvement involves four stakeholders, with different colors indicating their respective responsibilities.}
    \label{fig:methodology}
\end{figure*}

\begin{figure*}
    \centering
    \includegraphics[width=\linewidth]{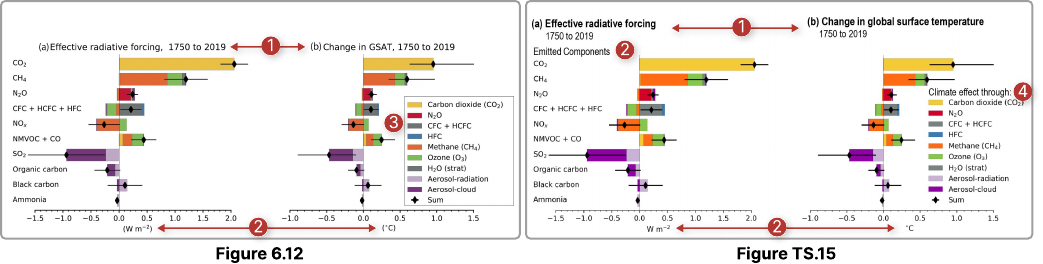}
    \caption{An example of an improvement in the IPCC report: (a) \href{https://www.ipcc.ch/report/ar6/wg1/figures/chapter-6/figure-6-12}{Figure 6.12} from IPCC Sixth Assessment Report, Working Group 1, the Full Report, aimed at scientists  \textcopyright~IPCC; (b) \href{https://www.ipcc.ch/report/ar6/wg1/figures/technical-summary/figure-ts-15}{Figure TS.15} from IPCC Sixth Assessment Report, Working Group 1, the Technical Summary, aimed at policymakers \textcopyright~IPCC.}
    \label{fig:example1}
\end{figure*}

We take \autoref{fig:example1} as an example to explain the iterative design process. The initial figure (left) is part of the IPCC report, \href{https://www.ipcc.ch/report/ar6/wg1/figures/chapter-6/figure-6-12/}{Figure~6.12}. It has been simplified by improvers working for the IPCC to appear in \href{https://www.ipcc.ch/report/ar6/wg1/figures/technical-summary/figure-ts-15/}{Figure~TS.15} of the technical summary for policymakers in collaboration with climate scientists using a process internal to the IPCC. 
We aim at a similar process outside of IPCC, designed to produce multiple solutions, potentially produced by a wide variety of improvers and validated by domain specialists.

\subsection{Setup and Initial State}\label{sec:initial_state}
Three elements must be prepared for the setup:
\begin{itemize}
    \item \textbf{Figure X}: the figure is selected by the \improvers for improvement, aimed at the \public.
    \item \textbf{Valid source code and data}: these are provided by \IPCC to produce Figure X. With code, the reproduced figure can be easily transformed into different formats, such as SVG, PDF, and PNG.
    \item \textbf{Learning objectives and assessments}: we adapt the learning-objective framework~\cite{Robbins2022Learning, Adar2021Communicative} to measure the achievement of the communication goals. \IPCC are those who understand the communication goals of the figure; therefore, they should be the ones to provide the communication goals in the form of learning objectives to guide the revision. Following the format from previous work~\cite{Adar2021Communicative},  
    A learning objective of an IPCC figure can be expressed in the format: ``\texttt{The viewer will [verb] [noun]}''. In our case, an example learning objective can be ``the \public will recall that climate warming now exceeds 1°C.''\\
    \IPCC also needs to provide one or more assessment tests for each objective. These assessments can take various formats, such as constructed responses or open-ended questions, as well as multiple-choice formats~\cite{Robbins2022Learning}. 
    \IPCC also needs to establish a scoring rule to calculate the final score considering all assessments. This rule may consider factors such as accuracy, response time, and proxies for engagement, including qualitative questions, among others.
\end{itemize}

The initial state is used to obtain the starting score for Figure X. 
In the initial state, \IPCC recruits an \textbf{Audience} from the \public to complete the assessment tests. The results, evaluated under the scoring rule, are denoted as $\textit{Score}_0$.
For example, \autoref{fig:example1} (left) is a figure representing the initial state.

\subsection{Iterations}\label{sec:iteration}
For each iteration, the \improvers and the \reviewers are the major stakeholders involved.
We introduce this section from their perspective.

\textbf{What do \improvers do?}
Improvers need to iteratively design improved figures based on Figure X. 
Each improved figure, then, needs to be evaluated through assessment tests.
Then, improvers will select the figure with the highest score $\textit{Score}_1$ and send this top-scoring figure, plus all related materials (including operation and assessment details), to reviewers for validation.

\textbf{What do \reviewers do?}
Two groups of experts are involved as reviewers: climate change experts and visualization experts.
The climate change experts serve as the primary reviewers, possessing knowledge of both climate change and basic visualization coding.
Since their coding expertise might be limited, they can seek assistance from visualization experts when necessary. 
If the climate change experts can make a decision independently, they can complete the validation on their own.
After their review, they will determine whether the improvement is valid. If it is deemed valid, the iteration is complete. If not, the improvers need to revise the improvement until reviewers consider it valid.

\textbf{How do \reviewers validate the figure?}
Reviewers need to validate two aspects: first, the figure itself, and second, the assessments for the figure.
We first discuss the validation for figures.

The primary principle for validating figures is that the visual strategy remains faithful to the original intent of the visualization. 
The data's visual representation must accurately reflect the original information without any undesired distortion.
Based on this principle and considering the characteristics of the IPCC, we have developed several key criteria for reviewers.
These criteria are intended to be broadly applicable, with the flexibility to be tailored to specific contexts as needed:

\begin{enumerate}[label=\textbf{C\arabic*}]
    \item \textbf{Data remains unchanged:}
    Data must remain unchanged during the analysis and visualization processes. Necessary transformations, such as applying a log scale, should be carefully applied to avoid introducing any bias or altering the original meaning of the data. This is crucial for ensuring the reliability and accuracy of conclusions, aligning with the FAIR principles~\cite{wilkinson2016fair}, and preventing misleading outcomes.
    \item \textbf{All changes must be transparent and valid:} 
    changes can include \textit{adding}, \textit{removing}, or \textit{modifying} elements, and all types of changes are acceptable as long as they are clearly visible and faithful to the original visualization intents. \textit{Adding} involves including new elements, such as annotation text. \textit{Removing} means taking out existing elements, such as a legend box. \textit{Modifying} involves altering elements, such as changing the color, shape, or text.
    A change is \textbf{transparent} when it is clearly documented, visible, and traceable, which enables stakeholders to understand what was changed. A change is considered \textbf{valid} when it preserves the original communication intent of the figure and is approved by domain experts. Together, transparency and validity are essential to prevent misunderstandings and to maintain trust in the visualization, ensuring that all stakeholders, including policymakers, the target audience, and scientists, can rely on the figure’s integrity.
\end{enumerate}

While adhering to these validation criteria does not guarantee the quality of the improvements, it provides the foundational framework for each iteration; all improvements must align with these criteria in our iterative process. 

\textbf{How do \improvers make improvements?}
It is the improvements that ultimately ensure the quality of the figures.
We assume that a single improvement consists of multiple operations. Since all operations must be validated, each operation must adhere to criteria C1 and C2. Describing all changes broadly as a single operation, such as ``title, axis, legend, and size are modified,'' can make it challenging for reviewers to validate the individual operations. 
Therefore, we provide instructions for documenting operations: each operation should focus on only one specific aspect. This approach makes it easier to reference and verify whether the modifications are valid.

We classified these aspects according to the Vega~\cite{satyanarayan2014declarative} and Vega-Lite~\cite{satyanarayan2016vega} specification of the Grammar of Graphics~\cite{wilkinson2012grammar}.
Our classification uses two levels. 
The first level includes size, axes, legends, titles, and marks. 
In the second level, for size, we consider width and height.
For axes, we consider elements such as labels, ticks, marks, titles, and grids. 
For legends, we include general settings, gradients, labels, symbols, symbol layout, and titles. 
Titles do not require further breakdown, and for marks, we refer to different types like bars, lines, and arcs. 
We provide an overview of the properties here; refer to the Vega-Lite documentation for detailed information on specific properties. 
Each operation focuses on a single component type (e.g., legend-label) to ensure clarity and precision.

\textbf{How do \improvers conduct assessments?} 
As we discussed before, the assessments were originally provided by IPCC.
However, for each improvement, the assessments will be slightly modified to accommodate the phrasing in the new figure.
For example, in the improved figure, some abbreviations will be expanded to their full names for better understanding.
In this case, the new question must reflect this change: if the question description includes any abbreviations, they should be replaced with the full words.
Using the new assessments, improvers can recruit appropriate target audiences as users and conduct the study. Based on the results and scoring rules, improvers can then calculate the final score for the improved figure.

\textbf{What materials do \improvers send to \reviewers?} 
After improvers select the top-score improvement, all materials related to a certain improvement need to be compiled and sent to reviewers. 
We give a list here for illustration.
\begin{itemize}
    \item Figure Information: figure number, iteration version number, and creation time.
    \item Author Information: name and email.
    \item Operation Information: detailed descriptions of each operation's modifications and what has been changed.
    \item Assessments Information: all the questions for the assessments, along with responses from various audiences, should be provided. Furthermore, the final score and the calculation method should also be documented.
\end{itemize}

\subsection{Iteration: Example}
\label{sec:iteration_example}
We use an example from the IPCC Sixth Assessment Report, Working Group 1, to help illustrate, as shown in \autoref{fig:example1}.
These two figures use the same data and visualization but appear in different reports. The left figure (\href{https://www.ipcc.ch/report/ar6/wg1/figures/chapter-6/figure-6-12/}{Figure~6.12}) is the original from the full report intended for scientists. The right figure (\href{https://www.ipcc.ch/report/ar6/wg1/figures/technical-summary/figure-ts-15/}{Figure~TS.15}) represents an improved version from the Technical Summary, designed for policymakers who may not have extensive expertise in climate change.
In this example, we consider Figure TS.15 as one improvement in one iteration. 
The improvement between the two figures can be observed in the title, axis, and legend, and in their sizes. 

\subsubsection{Operations}
In this case, there are five main operations. Each operation is numbered to correspond with the figure labels, with Op1 aligning with label 1 in the figure.
\begin{enumerate}[label=\textbf{Op\arabic*}]
\item Title modification: In figure TS.15b, the title is changed from ``Change in GSAT'' to ``Change in global surface temperature'' to avoid using the term ``GSAT,'' which policymakers may not be familiar with.
\item Axes modification: In figure TS.15a, the title ``Emitted Components'' was added above the y-axis. Additionally, for both the x-axes in figures TS.15a and TS.15b, the parentheses were removed, changing $(Wm_{-2})$ to $Wm_{-2}$ and $(^\circ\mathrm{C})$ to $^\circ\mathrm{C}$.
\item Legend general modification: In Figure TS.15, the legend was enclosed within a black-stroke box.
\item Legend title modification: The phrase ``Climate effect through'' was added above the legend in Figure TS.15.
\item Size modification: Since TS.15 is higher than Figure 6.12, the height needs to be adjusted to display all elements properly.
\end{enumerate}

\subsubsection{Implement Instructions}
\label{sec:iteration_detail}

According to C2, the process must be transparent, and \improvers should not be required to perform complex or time-consuming tasks to achieve this goal. 
Since the figures in IPCC reports are generated by code managed through Git for Working Group 1, we advocate for leveraging Git to its full potential to achieve these goals. 
Git, as a distributed version control system, facilitates efficient collaboration among developers by tracking changes, managing versions, and coordinating work across different branches of a project. 
Among Git’s various commands, \texttt{git diff} is particularly valuable as it compares file versions to highlight changes between commits, branches, or files, simplifying code review and understanding of modifications. 
This is a feature that can allow \reviewers to quickly identify the specific changes made.

We use the example to show how \improvers and \reviewers collaborate effectively using Git. This method is intended as a reference based on our current process, and we encourage users to adapt and improve it for more effective validation.
Future usage can follow our instructions or develop new validation methods, as long as criteria C1-C2 are met.

For each implementation, \improvers should first create a new branch, such as \texttt{``iteration1-improvement1''}.
Within a branch, each operation should be represented by a separate commit. 
As discussed in \autoref{sec:iteration}, each operation must be simple and easy to validate, so each commit should adhere to this principle. 
We recommend using the following format for commit messages: \texttt{[operation classification: specific modification]}.
For example, the commit messages for Operation in \autoref{fig:example1} could be:
\begin{enumerate}[label=\textbf{Op\arabic*}]
    \item \texttt{[title: update title text]}
    \item \texttt{[axes-title: add y-axis title, remove brackets in x-axis title]}
    \item \texttt{[legend-general: remove black stroke]}
    \item \texttt{[legend-title: add title "Climate effect through"]}
    \item \texttt{[size-height: increase height]}
\end{enumerate}

For how to implement the changes, \improvers can modify either the code or the graphics file. Since figures are generated by code, they can easily be exported into various formats, such as PDF, SVG, and PNG. 
We recommend using SVG~\cite{SVG}
because they are vector-based and use Extensible Markup Language (XML).

In cases where the \improvers need to modify the code, both the code file and the resulting graphic file will be affected. 
For some changes, the \improvers can also directly modify the SVG file, in which case only the graphic file will be affected.
For example, in \autoref{fig:example1}, Op1 can be implemented either in code or directly in the SVG. In Op2, the addition can be done via code, while the removal can be performed in either code or the SVG. Op3 can also be carried out in both code and SVG, whereas Op4 is implemented via code. Op5 is implemented via code. \improvers can choose the method they are familiar with. 
In both cases, the changes to the SVG files are detectable on GitHub through code diffs and visual diffs. For visual comparison of image files, it provides three methods: 2-up, swipe, and onion skin\footnote{\href{https://docs.github.com/en/repositories/working-with-files/using-files/working-with-non-code-files/}{docs.github.com/en/repositories/working-with-files/using-files/working-with-non-code-files}}. 

For each version of the improvement, the assessments will be slightly modified to accommodate the phrasing in the new figure.
For instance, in \autoref{fig:example1} (left), which represents the initial state, the question reads ``What emitted component has the largest absolute negative contribution on change in GSAT between 1750 and 2019?''
In \autoref{fig:example1} (right), which represents an improvement in iteration 1, the question would be ``What emitted component has the largest absolute negative contribution on change in global surface temperature between 1750 and 2019?''

In each iteration, \improvers can create multiple branches to correspond with various improvements, such as \texttt{``iteration1\-improvement2.''}
After assessing each improvement, they would select one and submit it to the \reviewers. 

It is also important to consider the format of the submitted file to ensure that the entire process remains efficient and effective.
We suggest using a combination of a detailed report to document all the necessary information, along with Git pull requests.
While the report should include all necessary details described in the last section, the pull request remains a crucial tool for \reviewers to interact with the changes.
Using the report and the interactive Git diff, \reviewers can verify whether the operations meet the two criteria (C1-C2).
The swipe and onion skin methods are particularly useful in our context. Since each commit contains a specific type of information, \reviewers can focus on a particular area using the commit message as a guide and visually compare the changes using these two methods.
\section{Case Study}\label{sec:case}
To demonstrate the practical application of our methodology, we conduct a case study using a figure from the IPCC. 
We selected Figure SPM.1 from the \href{https://www.ipcc.ch/report/ar6/wg1/figures/}{Sixth Assessment Report (AR6, 2021--2023) of Working Group I (WG1)}~\cite{fullreport2021}, as figures in the \href{https://www.ipcc.ch/report/ar6/wg1/figures/summary-for-policymakers/}{Summary for Policymakers} are specifically designed for non-expert policymakers and offer an ideal testbed for evaluating improvements.
To facilitate improvements, we required access to both the source code and data, as pixel-based images are difficult to modify.  
Since only the data for SPM.1 is available, we recreated a similar version using Python’s \texttt{matplotlib} and exported it in SVG format for easier manipulation.  
We discuss the challenges of this reproduction process in detail in \autoref{sec:reproduce}.

\begin{figure*}
    \centering
    \includegraphics[width=1\linewidth]{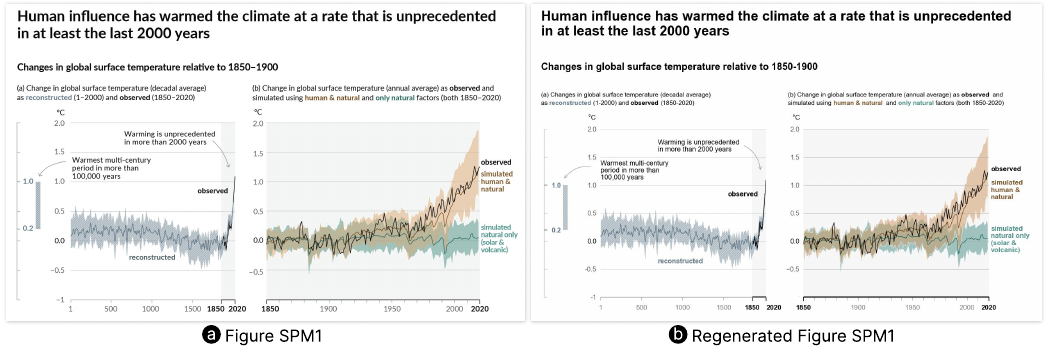}
    \caption{The \href{https://www.ipcc.ch/report/ar6/wg1/figures/summary-for-policymakers/figure-spm-1}{Figure SPM.1} used in our case study: (a) the original version from the Summary for Policymakers report, and (b) the regenerated version created for our study. \href{https://www.ipcc.ch/report/ar6/wg1/figures/summary-for-policymakers/figure-spm-1}{Figure SPM.1} is from the Summary for Policymakers of the IPCC Sixth Assessment Report, Working Group I \textcopyright~IPCC.}
    \label{fig:spm1}
\end{figure*}

In this case, we use a constructed example to demonstrate the full improvement workflow.
For illustration, we temporarily carry out the steps associated with certain roles (such as \improvers{} and visualization \reviewers{}), so that the process can be clearly understood by real stakeholders.

\subsection{Learning Objectives and Assessment}
\label{sec:lo-assessment}
This assessment is conducted both in the initial phase and throughout all iterations.
Typically, \IPCC determine the focus of the test, as they understand the key information each figure should convey. 
In this case, we (the authors) took on the role of \IPCC to demonstrate the process.
We further discussed the challenges we met during the process in  \autoref{sec:challenge-lo-assessment}. 

\subsubsection{Learning Objectives (LOs).}  
We acted in the role of \IPCC to derive the initial set of learning objectives for the figure.  
Since we were not the figure’s original designers or authors, we broadened our approach and built a thorough list with input from several groups of participants: initially a climate scientist, followed by three authors, and ultimately a group of HCI master’s students.

For the climate scientist, we reached out through our network and eventually identified a senior atmospheric scientist with over 20 years of experience, and held a one-hour discussion with him. During the meeting, we introduced our project and explained the concept of learning objectives, which follows the format: ``The Target Public will [verb] [noun].''
He read the figure and provided 2 LOs for chart (a) and 2 LOs for chart (b) via email.

Then, all three authors independently wrote learning objectives and corresponding assessment questions based on our own understanding of the figure and insights from our earlier discussion with the climate expert. 
We finalized 14 LOs for chart (a) and 11 for chart (b).

To further improve coverage, we involved a class of HCI master's students at one university. One of the authors gave a lecture introducing the project and the concept of learning objectives, and assigned homework that replicated our task: writing learning objectives and corresponding assessment questions for the same figure.
As a result, we collected 20 student submissions, yielding 99 LOs for chart (a) and 96 for (b). 
Due to the large volume and potential redundancies, we used GPT-4o to assist in organizing, clustering, and deduplicating the submissions. We interacted with ChatGPT step by step and manually verified its output at each stage.

This process resulted in a refined and more comprehensive list of learning objectives (11 for (a) and 11 for (b)), which better reflects both expert guidance and diverse audience interpretations.

\subsubsection{Assessment Design.}  
To assess our improvement, we designed a questionnaire to evaluate participants’ comprehension of the figure.
The questionnaire was derived from multiple sources. 
First, in identifying learning objectives, three authors and a group of HCI students concurrently proposed corresponding questions for each. 
From the 107 questions for (a) and 101 for (b), we collected (some objectives lacked corresponding questions), and we retained those that aligned with the finalized learning objectives.
To complement these, we adopted an LLM-assisted method inspired by VisQuestions~\cite{guan2025visquestions}, using GPT-4o to generate additional questions (3-4 questions for each LO).
All candidate items were grouped under their respective learning objectives, with redundant or overlapping questions removed. 
In some cases, simpler questions were merged into more challenging ones to increase difficulty. 
Moreover, when one learning objective was associated with only a single, overly straightforward question, we incorporated it into the options of related learning objectives rather than keeping it as a standalone item.
This iterative process produced a refined questionnaire that balanced breadth, depth, and difficulty.

Finally, we had 11 questions for chart (a) and 11 for chart (b), totaling 22 items. 
Among these, 9 were adapted into pre-test questions to assess participants’ baseline knowledge of climate change by removing figure-specific elements (see \autoref{sec:pre_test} in details).
They examine the same knowledge as the corresponding question in the formal test, but omit references to the figure.
These pre-test questions were placed before the presentation of the figures and tests. 
We refer to them as the \textbf{pre-test (9 items)} and the \textbf{formal test (22 items)}. 
Each question offers three answer choices plus an ``I don't know'' option, for which participants had to provide a brief explanation.
An additional open-ended question was included to collect participant feedback. 
The complete questionnaire is provided in the supplementary materials.  
We recorded the answer and response time for each question.

\subsubsection{Procedure.}
The evaluation for the initial state and one iteration was approved by the IRB from [redacted].
We recruited 100 participants from Prolific for both the initial state and the iteration, yielding a total of 200 participants.
Participants were fluent native English speakers and of legal age. They were compensated at a rate of €12 per hour. 

Each participant first completed a consent form and a short demographic survey, including age, education level, and familiarity with climate change and data visualization. Participants then completed 9 pre-test questions on climate change. They were then presented with chart (a) and chart (b), along with their captions and the corresponding questions. Participants answered 11 questions for chart (a) and another 11 for chart (b), separated by a short break. They were allowed to take as much time as they needed to examine the charts carefully. 

\subsection{Initial State}
\label{sec:case_initial}

The initial state is implemented with our reproduced figure as shown in \autoref{fig:spm1}(b). We call it \textbf{V0} in the following sections.
We recruited 119 participants in total; 100 completed all tasks and were analyzed. 17 dropped out, and 2 were excluded due to unrealistically short completion times (under 6 s) combined with low accuracy or inconsistent responses.
Participants ranged in age from 18 to over 65 (Mean $\approx 41$, Median $\approx 39$), with the majority (54\%) between 25 and 44 years.
Education levels were high school (29), bachelor’s (48), master’s (16), PhD (3), and other (4). 
Self-reported climate expertise comprised very familiar (18), familiar (42), and somewhat familiar (40). 
Visualization frequency was daily (8), several times a week (39), once a week (11), several times a month (23), once a month (17), and other (2).
The median response time was 31 minutes 9 seconds.

\begin{figure}[htbp]
    \centering
    \includegraphics[width=\linewidth]{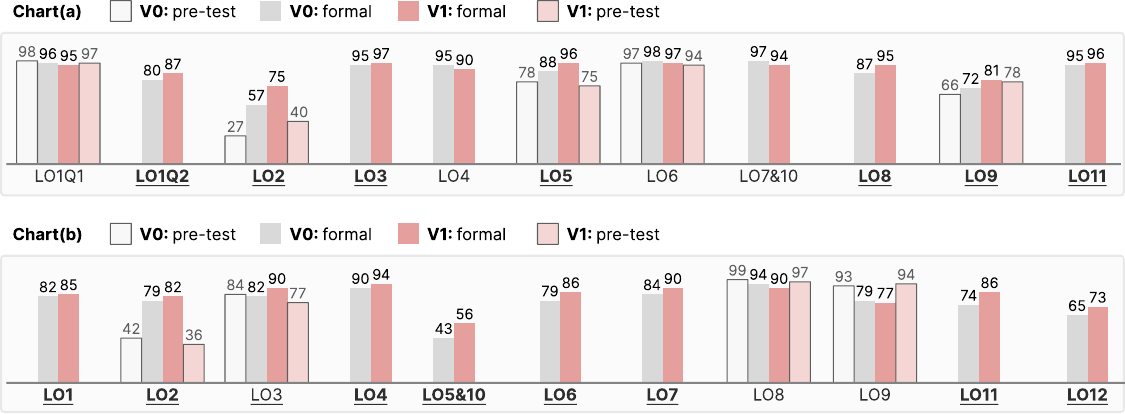}
    \caption{The overall accuracy of chart (a) and chart (b) in \textbf{V0} and \textbf{V1}. The x-axis labels indicate question indices guided by learning objectives (LOs). Some questions involve multiple LOs, which we connect using ``\&''.
    The bold underlined LOs indicate that V1 achieves higher accuracy than V0.}
    \label{fig:accuracy}
\end{figure}

\subsection{Iteration 1}

\begin{figure*}[htbp]
    \centering
    \includegraphics[width=\linewidth]{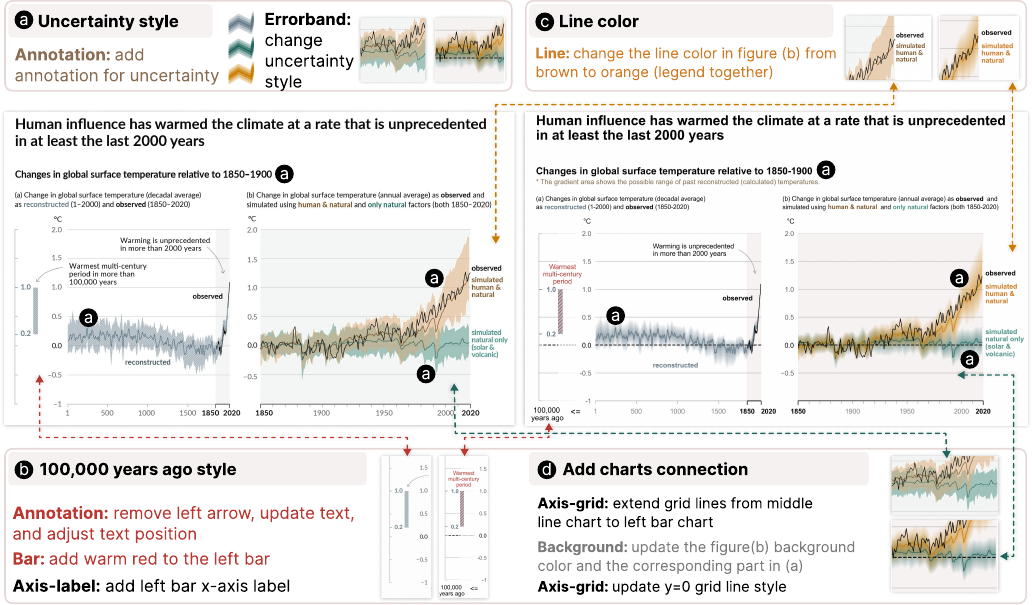}
    \caption{\yl{Our improvements to Figure SPM.1 are illustrated in (a)–(d); they highlight four directions.
    Each detailed modification is indicated by label (a) near elements or dashed arrows linking the text to the middle chart. \href{https://www.ipcc.ch/report/ar6/wg1/figures/summary-for-policymakers/figure-spm-1}{Figure SPM.1} is from the Summary for Policymakers of the IPCC Sixth Assessment Report, Working Group I \textcopyright~IPCC.}}
    \label{fig:spm1_improvement}
\end{figure*}

We served as \improvers in this iteration, refining the figure based on items with low accuracy as shown in \autoref{fig:accuracy}. 
Specifically, all authors iteratively discussed design options and refined ideas through multiple rounds of feedback.
All improvements were shown in \autoref{fig:spm1_improvement}, accompanied by the corresponding low-accuracy questions:
\begin{itemize}
    \item enhancing the display of uncertainty with gradient areas and prominent annotations (a: LO9, b: LO10); 
    \item improving the display of 100,000 years before through clearer annotation, a more prominent bar, and the addition of an x-axis label (a: LO2);
    \item adjusting color choices for better distinction for ``natural and human'' (b: LO2, b: LO5\&10);
    \item relating chart (a) and chart (b) with a subtle red-tinted background (b: LO11).
\end{itemize}
One author implemented the modifications in Python on GitHub, followed the operations discussed in \autoref{sec:iteration_example}, while the other authors served as \reviewers and validated the commits to ensure the process remained transparent and faithful. 
The texts below the gray boxes (a–d) of \autoref{fig:spm1_improvement} are the actual commit messages.

For this round of the study, we used the improved figure together with an adapted questionnaire, in which some formal test questions were modified to reflect the figure changes (e.g., the line color changed from brown to orange). 
This version is referred to as \textbf{V1} in the following sections (\autoref{fig:spm1_improvement}e).
We recruited a total of 120 participants; 100 completed all tasks and were included in the analysis, and 20 dropped out before finishing.
Participants ranged in age from 18 to over 65 (Mean $\approx 39$, Median $\approx 33$), with the majority (68\%) between 25 and 44 years.
Education levels included high school (27), bachelor’s (47), master’s (20), PhD (3), and other (3). 
Self-reported climate expertise was very familiar (17), familiar (44), somewhat familiar (34), and not familiar (5). 
Visualization frequency was reported as daily (4), several times a week (35), once a week (21), several times a month (22), once a month (16), and 'other' (2).
The median response time was 31 minutes and 51 seconds.

\subsection{Comparative Analysis of V0 and V1}
To compare participants' performance on the two chart versions, we applied a generalized linear mixed model (GLMM)~\cite{bolker2015linear}. 
Participants differ in prior knowledge, as measured by the pre-test, and items vary in difficulty, so a simple accuracy average would conflate version effects with these factors.
Because responses are binary (correct/incorrect), GLMM is particularly appropriate: it models binary outcomes while accounting for participant- and item-level variation, thereby isolating the chart version effect~\cite{dean2007generalized}.

For each response in the pre-test and formal test for charts (a) and (b) (33 items in total), 
we coded correct answers as $1$ and all other responses, including ``I don't know'', as $0$. 
We specified a logistic GLMM of the form:
\[
\logit\,\Pr(Y_{ij}=1) 
= \beta_0 
+ \beta_1 \text{V}_{ij} 
+ \beta_2 \text{P}_{i} 
+ \beta_3 (\text{V}_{ij} \times \text{P}_{i})
+ u_i + v_j
\]
where $Y_{ij}$ denotes the response of participant $i$ to item $j$, 
$\text{V}_{ij}$ is the chart version (V0 or V1), 
$\text{P}_{i}$ is the mean-centered pre-test score, 
and $u_i \sim \mathcal{N}(0, \sigma_u^2)$ and $v_j \sim \mathcal{N}(0, \sigma_v^2)$ are random intercepts for participants and items, respectively. 
We specified the following fixed effects in the model:
\begin{itemize}
    \item $\beta_0$: \textbf{intercept}, representing the baseline probability of a correct response (for an average participant answering an average item under V0).
    \item $\beta_1$: \textbf{chart version effect}, comparing V1 against V0.
    \item $\beta_2$: \textbf{pre-test effect}, capturing how prior knowledge (mean-centered pre-test score) predicts accuracy.
    \item $\beta_3$: \textbf{interaction effect}, testing whether the version difference depends on prior knowledge.
\end{itemize}

We fit the model to our data in R. The data and analysis scripts can be found in the supplementary materials.
\begin{itemize}
    \item \textbf{Intercept} ($\beta_0 = 2.153$, $p < .001$): 
    baseline accuracy under V0 for an average participant on an average item was about $89.6\%$.
    \item \textbf{Chart version effect} ($\beta_1$): significant 
    ($\beta = 0.53$, $p = .006$), corresponding to an odds ratio of 1.70. 
    Participants viewing V1 had 70\% higher odds of answering correctly (OR = 1.70), corresponding to an absolute increase of about four percentage points (from 89.6\% to 93.6\%).
    
    \item \textbf{Pre-test effect} ($\beta_2$): positive 
    ($\beta = 0.236$, $p = .020$), meaning that each additional pre-test point increased the odds of a correct response by about $27\%$.
    
    \item \textbf{Interaction effect} ($\beta_3$): not significant ($p = .54$) but still positive.
    The benefit of \textbf{V1} was consistent across participants regardless of prior knowledge, 
    indicating that both low- and high-knowledge participants benefited similarly from the improved chart.
\end{itemize}

The result indicate that the improved version \textbf{V1} significantly improved the participants' understanding of the visualization, independent of participants' prior knowledge. 
Therefore, it shows the effectiveness of the improvement in supporting the achievement the learning objectives of the visualization. 
Combining this result with \reviewers's work on monitoring the improvement process, our case shows an example of using our proposed methodology to achieve an effective and faithful improvement of an IPCC figure.
     \section{Challenges Beyond Expectations}

To inform future work, we recorded the difficulties we encountered and our solutions, especially in steps we initially assumed would be easy.

\subsection{Reproducing IPCC Figures}\label{sec:reproduce}
During the whole process, we made the assumption that \emph{the figures in the report are reproducible}.
As mentioned in \autoref{sec:initial_state}, the valid source code and data to generate Figure X in SVG vector format is one requirement for the initial state. 
This is essential because the figure often needs to be shared and modified by multiple people.

\subsubsection{Code and Data not Available} 
For the \href{https://www.ipcc.ch/report/ar6/wg1/figures/}{Sixth Assessment Report (AR6, 2021--2023) of Working Group I (WG1)}~\cite{fullreport2021}, the IPCC has allegedly provided the source code\footnote{\href{https://github.com/ipcc/}{github.com/ipcc/}} and the data\footnote{\href{https://data.ceda.ac.uk/badc/ar6_wg1/data}{data.ceda.ac.uk/badc/ar6\_wg1/data}} for the charts through separate public repositories. 
However, the source code and datasets are not only stored separately (on GitHub and CEDA, respectively), making them difficult to locate, but also incomplete (i.e., not all figures have their associated data and code available).
Therefore, we encountered four typical cases when we reproduced IPCC figures, as also highlighted by~\cite{ying:hal-04744236}:

\begin{itemize}
    \item \textbf{Neither code nor data is available.}  
    In these cases, reproduction is entirely infeasible. Without either component, we could not attempt any reconstruction.

    \item \textbf{Code is available, but data is missing.}  
    We were able to examine the implementation logic, but could not execute the scripts. For some figures, we contacted the authors or the IPCC to request missing datasets, and managed to retrieve part of them, enabling partial or full reproduction.

    \item \textbf{Data is available, but code is missing.}  
    This situation allowed us to explore the data, but the lack of code forced us to reconstruct or approximate the original visualization logic. 
    These reconstructions could not replicate the original figures exactly and were only approximated to the best of our ability. Our case study discussed in \autoref{sec:case} is this scenario. 

    \item \textbf{Both code and data are available.}  
    Even in the best-case scenario, direct reproduction was not always successful due to broken dependencies, hardcoded paths, or outdated environments. We reconfigured several environments with updated Python versions and packages, and successfully reproduced some figures.
\end{itemize}

\begin{figure*}[htbp]
    \centering
    \includegraphics[width=1\linewidth]{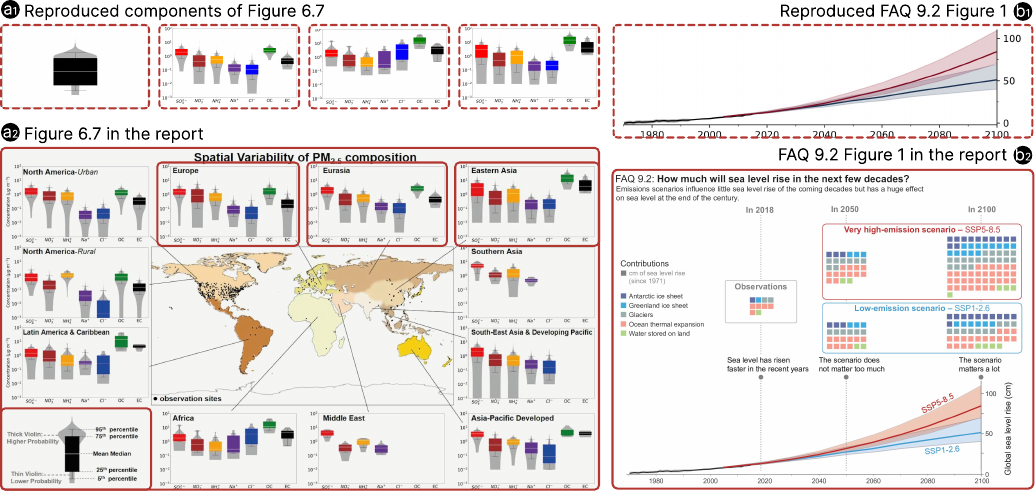}
    \caption{Two conditions are encountered during the reproduction of \href{https://www.ipcc.ch/report/ar6/wg1/figures/chapter-6/figure-6-7}{Figure 6.7}: (a1-a2) the generated figure is incomplete, and (b) the generated figure has been modified. \href{https://www.ipcc.ch/report/ar6/wg1/figures/chapter-6/figure-6-7}{Figure 6.7} is from IPCC Sixth Assessment Report, Working Group 1\textcopyright~IPCC.}
    \label{fig:reproduce}
\end{figure*}

\subsubsection{Mismatch Between Reproduced and Original Figures} 
We also encountered another challenge: even when the code ran, the reproduced figure and the figure in the IPCC report were not identical. These figures required post-processing. Below, we list some of the differences we have identified:
\begin{enumerate}
\item Many generated charts are actually components of larger figures rather than complete ones. This includes sub-figures in most conditions and also includes some other conditions where charts and legends are generated separately and then combined. An example can be seen in \autoref{fig:reproduce}(a), where 13 components are generated for the figure. Some of these components are displayed in \autoref{fig:reproduce}(a1). We need to manually combine these code-generated components to get the original larger figure.
\item There are minor style differences compared to the original figures, such as the presence of outer boxes around legends, variations in color usage, and the inclusion or exclusion of legends or axis labels. These repositories likely contain versions close to, but not exactly, the final printed versions. In particular, for color variation, although we’re using similar color schemes, the colors are not identical. This discrepancy might be attributed to variations in printing or compiling methods.
\item Some figures are not classic charts that can be generated by code. Like figures in summary reports for policymakers and frequently asked questions, some are infographics. In these cases, the code-generated figures serve as a starting point, but improvers may add decorations and annotations and make substantial modifications to the final figures. \autoref{fig:reproduce}(b) provides an example where the basic outline of the line chart from \autoref{fig:reproduce}(b1) is retained, but the overall style has been completely altered.
\end{enumerate}

In our case study (\autoref{sec:case}), since the data was available but the code was missing, we wrote our own Python code to reproduce the figure based on the image.
While we made our best effort to replicate the figure, the result is not identical to the original. However, the discrepancies are minor style differences; therefore, they should have very little effect on the understanding of the visualization based on our knowledge.

\subsection{Deriving Learning Objectives and Assessment}
\label{sec:challenge-lo-assessment}

\subsubsection{Constructing Learning Objectives in the Absence of \IPCC}

Ideally, the learning objectives should be provided by the \IPCC, as they are most familiar with the intended messages of each figure and the report as a whole. However, in our case, we were unable to obtain these objectives directly.

This constraint would not be uncommon. Given the global significance of climate change, IPCC figures are likely to attract increasing attention. Many individuals may find themselves in a situation where they want to improve an IPCC figure. For example, a student may want to redesign a figure to better help their peers understand climate-related information. As a result, \IPCC may not be able to promptly respond to every request for the learning objectives. Furthermore, these objectives are not currently accessible to the public.

Our approach to tackling this challenge could be a valuable reference for future IPCC figure improvement efforts. 
To ensure that the learning objectives we constructed were both comprehensive and valid, we adopted a rigorous learning objective development process (see \autoref{sec:lo-assessment}). The central idea of our approach was the combination of expertise from both climate science and data visualization, because IPCC figures are visualizations with climate science content. Specifically, we consulted with climate scientists, incorporated input from HCI master’s students, and drew upon our own expertise as visualization researchers. In this way, we aimed to compensate for the absence of officially learning objectives from the \IPCC.

\paragraph{Pre-test Considerations}
\label{sec:pre_test}
In addition to the formal questions derived from the learning objectives, we included a pre-test to capture participants’ prior knowledge of climate change. 
This decision was partly shaped by our study setting: given the limited sample size, we could not rely on randomization alone to balance prior knowledge across conditions. We initially explored existing instruments such as those from Yale University~\cite{gustafson2025yale} and NASA~\cite{nasa_quiz}, but found that their broad coverage of climate change concepts did not align well with the specific knowledge targeted in our tasks. 
To address this, we constructed the pre-test directly from our own question set, omitting visual references in order to focus on underlying concepts. 
One author drafted the items, and three others reviewed them to ensure that the assessed knowledge points were comparable to those in the formal test. 
Nevertheless, achieving strict one-to-one correspondence proved challenging: the multiple-choice options in the formal test were tightly coupled to specific graphical elements, and once the figures were removed, it became difficult to maintain equivalent difficulty and to preserve exactly the same knowledge focus.
This consideration further motivated our use of a GLMM, which models variation at both the participant and item levels without requiring direct alignment between pre- and post-test questions.

This challenge was also reflected in our analysis: for instance, accuracy on LO2 (Chart (b)) declined from the pre-test to the formal test across both versions, as shown in \autoref{fig:accuracy}. 
The pre-test only required recalling the main driver of recent warming (``human activities''), while the formal question demanded evaluating chart-based causal claims. 
Many participants selected the partially correct but ultimately misleading option (B), highlighting the cognitive challenge of integrating prior knowledge with chart-based evidence.

\section{Discussion}
We discuss the broader implications of our methodology, focusing on its adaptability across audiences and institutions, the tension between trust and flexibility, and its long-term impact. We also acknowledge limitations and outline directions for future work.

\subsection{Adapting Figures for Diverse Audiences and Institutions}
In real scenarios, a single version of a figure cannot serve all audiences. The improvement in our case study primarily targets the general public, but groups such as teenagers and elders differ substantially in prior knowledge, chart literacy, and information needs. 
For instance, teenagers are both highly affected by climate change and influential in shaping future awareness, yet they often have limited chart literacy. For this group, improvements should begin by adapting learning objectives, for example, by emphasizing overall trends, remove some elements, and avoiding jargon. 
Policymakers would require additional information not specified by the learning objectives, such as an executive summary figure showing trends and risks that are clearly actionable to them. Scientists, on the other hand, would want all the details, including uncertainty and specialized information such as short model names.

Beyond individual audiences, the same methodology can also be adopted by other science-policy bodies, such as \href{https://www.ipbes.net/}{IPBES} and \href{https://www.who.int/}{WHO}, who face similar challenges in communicating complex findings to heterogeneous stakeholders. This demonstrates the broader applicability of our framework across domains where scientific rigor must be balanced with effective public communication, offering a generalizable, scientifically verifiable process for figure improvement.

\subsection{The Tension Between Trust and Flexibility}

We found that collaboration between \improvers and \reviewers fundamentally lies in balancing trust and flexibility. This tension is not confined to a single aspect but often manifests in multiple dimensions.

For the figure improvement tool, we currently require \improvers to make modifications either by programming or directly in the SVG code. This ensures a clear and transparent process, as all changes can be tracked via e.g., \texttt{git diff}. However, these coding-based operations are not what designers are most familiar with. They may be more comfortable using tools like Figma or Adobe Illustrator, which are easier for them to work with when making modifications. 
These changes, although they can be viewed via visual diffs on GitHub cannot be tracked.
This creates a clear tension. Code-based operations provide full trace but lower flexibility, while mainstream design tools provide high flexibility but lower trust, since changes are difficult to trace. 

Beyond tools, a similar tension arises when deciding whether improvement operations should involve removing elements tied to learning objectives, particularly those concerning uncertainty.
For climate scientists, uncertainty is a non-negotiable element of rigor and trust, yet for the general public it often causes confusion and hampers comprehension~\cite{zehr2000public}. 
Removing such objectives is technically supported within our methodology and the comparison using GLMM, allowing results to be recomputed consistently from the remaining question pool.
From a flexibility standpoint, this makes figures more audience-appropriate. However, \reviewers, especially climate scientists, may resist such operations, seeing them as a compromise of scientific fidelity. 
This highlights a second trust–flexibility tension: should transparent documentation of removed objectives be enough to maintain credibility, or must some objectives always remain intact for all audiences?

\begin{figure}[htp]
    \centering
    \includegraphics{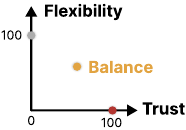}
    \caption{A balanced coordination between flexibility and trust for effective collaboration between \improvers and \reviewers.}
    \Description{A simple 2D diagram illustrating the coordination between flexibility and trust. The x-axis represents Trust, ranging from 0 to 100, with a red dot placed at 100. The y-axis represents Flexibility, also ranging from 0 to 100, with a gray dot at 100. A yellow point labeled Balance appears in the middle space between the two axes, symbolizing the desired balance between the two dimensions.}
    \label{fig:balance}
\end{figure}

As shown in \autoref{fig:balance}, we initially envisioned a middle ground (yellow point) where both trust and flexibility are optimized, but in practice, such a solution has not yet been realized.
How to design hybrid solutions that maintain transparency and accountability while offering \improvers the flexibility of improving figures is a direction for future research and tool development.

\subsection{Long-term Advantages and Future Impact}
Our methodology brings long-term advantages that extend beyond individual figure improvements.
First, when skeptics raise doubts, every modification can be backed by transparent records about improvements and tests, so that skeptical claims can be addressed with transparent evidence, providing a stronger foundation for public trust.
Second, by archiving all improvement iterations, the methodology creates a reusable body of knowledge. Future \improvers can build on prior efforts, turning isolated modifications into transferable design patterns. Over time, such documentation may foster a community of contributors that continuously refines figures for diverse audiences.

Together, these dynamics form a virtuous cycle: greater transparency and credibility lead to broader audience understanding; broader understanding motivates more contributors to tailor figures; and these contributions, in turn, further expand recognition and acceptance of climate change information. 
In this way, the methodology evolves from a tool for figure revision into a sustainable ecosystem that amplifies the communicative power of scientific data and knowledge.

\subsection{Limitations and Future Work}
While our study demonstrates the potential of the proposed methodology, several limitations remain. 
First, the procedures required in our methodology, even though necessary for the purpose, were relatively complex and time-consuming, which may have led to participant dropout.
This introduces a future avenue for more scalable or lightweight evaluation methods.
Second, as a preliminary exploration, our case study was conducted with a relatively small number of participants and validated only by us as visualization experts. Although our modifications were validated to ensure their methodological rigor, future work should expand to larger groups and include climate scientists as reviewers to ensure broader applicability and domain relevance.
Third, as mentioned in \autoref{sec:background}, we conducted close collaboration with IPCC staff with the goal of exploring ways to improve the communication of climate figures.
Through the case study, we show our methodology using one figure from IPCC Working Group I and verify the effectiveness of our proposed method, which demonstrates its potential for eventual integration into the IPCC official workflow or guidelines.
However, at this time, we could not guarantee a formal large-scale adoption, as implementation within an organization of IPCC’s scale requires multiple levels of review and approval.
Limitations and potential extensions also exist in terms of the output format, since our work focuses on a static, improved figure as output. However, the climate figures from IPCC could be communicated in animated video, GIF content, interactive web elements, or even in audio (for people with visual impairment). 
Further studies are needed to establish such relevant criteria for improvements and review in these contexts, while keeping the overall ``chain of trust'' idea at the core of this work.
\section{Conclusion}
We presented a constructive methodology for improving IPCC figures that emphasizes transparency, scientific rigor, and audience-centered communication. 
By grounding design revisions in official data and learning objectives, and by involving \improvers, \reviewers, \IPCC, and \public, our approach ensures that visual improvements enhance comprehension without compromising scientific rigor.
The case study demonstrates the feasibility of this workflow and highlights how iterative design and systematic evaluation can bridge the gap between scientific intent and public understanding. 
At the same time, our experience surfaces important challenges---such as reproducing complex figures and designing valid, scalable assessments—that future work should address to make the process more available and efficient. 
Beyond the IPCC, our methodology can inform other science–policy communication contexts, offering a pathway toward trustworthy, adaptable, and evidence-based visualization practices that better support diverse audiences in engaging with critical global issues.

\begin{acks}
We thank Lina Sitz and Clotilde Pean from the IPCC Technical Support Team for their support, Court Strong for his early feedback on the learning objectives, and
Elsie Lee-Robbins and Eytan Adar for their feedback on using learning objectives practically. 
\end{acks}

\bibliographystyle{ACM-Reference-Format}
\bibliography{IPCC_CHI25}

\end{document}